\documentclass[12pt]{article}

\usepackage{graphicx}
\begin{document}

\begin{center}
{\bf Holographic superconductor with nonlinear arcsin-electrodynamics } \\
\vspace{5mm} S. I. Kruglov
\footnote{E-mail: serguei.krouglov@utoronto.ca}
\underline{}
\vspace{3mm}

\textit{Department of Chemical and Physical Sciences, University of Toronto,\\
3359 Mississauga Road North, Mississauga, Ontario L5L 1C6, Canada} \\
\vspace{5mm}
\end{center}

\begin{abstract}
We investigate holographic s-wave superconductors with nonlinear arcsin-electrodynamics
in the background of Schwarzschild anti-de Sitter black holes. The analytical Sturm-Liouville eigenvalue problem
is explored and we assume that the scalar and electromagnetic fields do not influence on the background
metric (the probe limit). The critical temperatures of phase transitions depending on the parameter of the model is obtained. We show that in our case the condensation formation becomes easier compared to Born-Infeld nonlinear electrodynamics. The critical exponent near the critical temperature is calculated which is 1/2. With the help of the matching method we derive analytic expressions for the condensation value and the critical temperature. The real and imaginary parts of the conductivity in our model, making use of an analytical method, are computed.
\end{abstract}

\section{Introduction}

The correspondence between anti-de Sitter (AdS) spacetime and a conformal field
theory (CFT) on a boundary, which goes from the string theory \cite{Maldacena}, \cite{Gubser}, \cite{Witten}, is now of great interest.
The conjecture proposed by Maldacena \cite{Maldacena} connects type-IIB string theory with N = 4 SU(N) supersymmetric
Yang-Mills theory which is a conformal field theory. This duality was named AdS/CFT correspondence. Then this conjecture was generalized to other gravitational backgrounds and without supersymmetry and conformal symmetry.
In other words the AdS/CFT correspondence is a strong/weak duality that gives the link between the strongly coupled quantum field theory and the dual weakly interacting gravitational theory. Thus, this holographic duality provides us a possibility to study strongly interacting systems so that a gauge invariant operator in CFT corresponds to a dynamical field in the bulk gravitational theory.
As a result, the infrared physics of the bulk gravitational theory near the boundary possesses a counterpart in the ultraviolet physics of dual CFT. This holographic correspondence allow us to study a strongly interacting field theory of condensed matter system in $d$ dimensions by a classical gravity in $d+1$ dimensions \cite{Hartnoll}, \cite{Hartnoll_1}, \cite{Herzog}.
This AdS/CFT correspondence can help to understand the mechanism of high temperature superconductors in condensed matter physics \cite{Hartnoll2}.
It was shown in \cite{Gubser_1}, \cite{Gubser_2} that if gravity is coupled to gauge fields then charged scalars can condense near the horizon
and black hole horizons lead to spontaneous breaking of an Abelian gauge symmetry, i.e. black holes can superconduct.
The thermodynamics of CFT is in agreement with the thermodynamics of the black hole in the dual gravitational theory. So the Hawking temperature of the black hole is identical with the temperature in the CFT and a chemical potential in this CFT corresponds to a gravity theory with a conserved charge. The thermal entropy of CFT equals the entropy of the black hole, which is the area of the black hole horizon over $4G_N$, and the free energy is connected with the Euclidian on-shell bulk action.

Different aspects of holographic superconductors were studied in \cite{Umeh}-\cite{Roychowdhury_0}.
A lot of works were done in the probe approximation when the scalar and electromagnetic fields do not effect on the metric and
back-reaction in the bulk gravitational theory is ignored.
In the work \cite{Siopsis} authors explored the Sturm-Liouville variational method to investigate
the properties of holographic superconductors. In the probe limit the critical temperature and critical exponent were studied
for s-wave, p-wave and d-wave superconductors \cite{Gangopadhyay}-\cite{Ge}.
In \cite{Roychowdhury_0}, \cite{Yao}, \cite{Lai} authors studied the holographic superconductors in the framework of Born-Infeld (BI) nonlinear electrodynamics.
It was shown that in this case the critical temperature decreases compared to the holographic superconductor with the linear Maxwell fields. Authors of \cite{Liu} investigated the process of non-equilibrium condensation in holographic superconductors with
BI and logarithmic electrodynamics. Holographic superconductors with BI, logarithmic and exponential nonlinear electrodynamics were studied also
in \cite{Jing_1}-\cite{Sheykhi}. Holographic superconductors with BI electrodynamics away from the probe limit were investigated in \cite{Ghoraia}.

In this paper, we study holographic s-wave superconductors with arcsin-electrodynamics proposed in \cite{Kruglov} (see also \cite{Kruglov_1}).
The attractive feature of this model of nonlinear electrodynamics is that the electric field of a point-like charge is finite at the origin and the static electric energy of a particle is also finite. Thus, this model is similar to BI electrodynamics. The regular black hole solution within arcsin-electrodynamics was obtained in \cite{Kruglov_2}.
We use here four-dimensional AdS$_4$ which can provide analogies to phenomena in thin-film superconductors.
The Sturm-Liouville analytical approach is explored and we concentrate on the probe limit. We use also the matching method to calculate the condensation values and the critical temperature. We compute the real and imaginary parts of the conductivity with the help of an analytical method.

The structure of the paper is as follows. In section 2, we write down the field equations in $4$ dimensions for AdS$_4$ black holes with arcsin-electrodynamics. The critical temperature is calculated as a function of the charge density in section 3 making use of the Sturm-Liouville analytical approach. In section 4, we obtain the condensates of the scalar operators. Analytic expressions for the condensation value and the critical temperature are obtained using the matching method in section 5. In section 6 we calculate the conductivity in our model with making use of an analytical method. We discuss the results obtained in section 7.

\section{Field equations}

Our goal is to investigate the formation of a scalar hair of AdS$_4$ black hole.
In order to construct a holographic s-wave superconductor in the probe limit, we consider
the background of the Schwarzschild-AdS$_4$ black hole with the line element
\begin{equation}
ds^2=-f(r)dt^2+\frac{1}{f(r)}dr^2+ds_2^2,
\label{1}
\end{equation}
where the “transverse” metric $ds_2^2$ can be the metric on a unit 2-sphere but we assume the flat (planar)
metric $ds_2^2$ on $\textbf{R}^2$ to be in the form of $ds_2^2=r^2(dx^2+dy^2$).
The metric function of the Schwarzschild-AdS$_4$ black hole is given by \cite{Hartnoll}
\begin{equation}
f(r)=\frac{r^2}{L^2}-\frac{M}{r},
\label{2}
\end{equation}
where $L$ is the curvature radius of the AdS$_4$ black hole and the mass of the black hole is $M/2$. The negative cosmological constant of AdS$_4$ spacetime $\Lambda$ is equal to $\Lambda=-3/L^2$. The asymptotically AdS$_4$ region occurs at large $r$ and a regular
black hole horizon radius $r_+$ is determined by roots of the equation $f(r_+)=0$. From Eq. (2) we obtain $r_+=(ML^2)^{1/3}$. The Hawking temperature is given by
\begin{equation}
T_H=\frac{\kappa}{2\pi}=\frac{f'(r_+)}{4\pi}=\frac{3M^{1/3}}{4\pi L^{4/3}},
\label{3}
\end{equation}
where $\kappa$ is the surface gravity. In the background defined by the metric function (2) we consider the Lagrangian density of
nonlinear arcsin-electrodynamics and a charged complex scalar field
\begin{equation}
{\cal L} ={\cal L}_0 -|(\nabla_\mu-iqA_\mu)\psi|^2-V(|\psi|),
 \label{4}
\end{equation}
where the Lagrangian density of arcsin-electrodynamics was proposed in \cite{Kruglov} and is
\begin{equation}
{\cal L}_0 = -\frac{1}{\beta}\arcsin(\beta{\cal F}),
 \label{5}
\end{equation}
where ${\cal F}=-(1/4)F^{\mu\nu}F_{\mu\nu}$ and $\beta$ is a dimensional parameter of the model. The potential function $V(|\psi|)$ of the scalar field $\psi$ can be chosen in different shape. The simplest form is \cite{Hartnoll}
\begin{equation}
V(|\psi|) =m^2|\psi|^2,
 \label{6}
\end{equation}
where the mass squared $m^2$ can be positive or negative (when the field $\psi$ is a tachyon). We consider the case \cite{Hartnoll} $m^2=-2/L^2$
which does not lead to instability and is above the Breitenlohner-Freedman bound \cite{Breitenlohner}. We assume that the field $\psi$
is small and do not back-react signiﬁcantly upon the geometry. If $\beta\rightarrow 0$ the Lagrangian density (5) is converted into the Maxwell Lagrangian density ${\cal L}_M=-{\cal F}$, i.e. the  correspondence principle holds. Our purpose is to study how the parameter $\beta$ influences on the condensate formation in the dual theory (CFT). We are going to obtain solutions of the scalar hair that do not back-react. From Eqs. (4)-(6) we obtain field equations as follows:
\begin{equation}
\frac{1}{\sqrt{-g}}\partial_\mu\left(\frac{\sqrt{-g}F^{\mu\nu}}{\sqrt{1-(\beta{\cal F})^2}}\right)=iq\left(\psi^*\partial^\nu\psi
-\psi\partial^\nu\psi^*\right)+2q^2A^\nu|\psi|^2,
 \label{7}
\end{equation}
\[
\partial_\mu\left(\sqrt{-g}\partial^\mu\psi\right)-iq\sqrt{-g}A^\mu\partial_\mu\psi-iq\partial_\mu(\sqrt{-g}A^\mu\psi)-q^2\sqrt{-g}A_\mu A^\mu \psi
\]
\begin{equation}
-\sqrt{-g}m^2\psi=0.
 \label{8}
\end{equation}
We assume that the Abelian gauge field $A_\mu$ and the scalar field $\psi$ have an ansatz \cite{Hartnoll}
\begin{equation}
A_\mu=(\phi(r),0,0,0)),~~~~\psi=\psi(r).
 \label{9}
\end{equation}
A convenient choice of units is $q=1$ and the phase of the scalar field be zero, i.e. the field $\psi$ is real. Making use of Eqs. (7), (8) and (9) one obtains
\begin{equation}
\frac{1}{r^2}\partial_r\left(\frac{r^2\phi'(r)}{\sqrt{1-\beta^2\phi^{'4}(r)/4}}\right)-
\frac{2\psi^2(r)}{f(r)}\phi(r)=0,
 \label{10}
\end{equation}
\begin{equation}
\psi''(r)+\left(\frac{2}{r}+\frac{f'(r)}{f(r)}\right)\psi'(r)+\frac{\phi^2(r)}{f^2(r)}\psi(r)+\frac{2}{L^2f(r)}\psi(r)=0.
 \label{11}
\end{equation}
Eq. (11) is the same as for the case of linear Maxwell's equations. At $\beta\rightarrow 0$ we come to classical electrodynamics and Eq. (10) is converted to one investigated in \cite{Hartnoll}. For simplicity we use units with $L=1$ ($r_+=M^{1/3}$). At the horizon, $r=r_+$, $\phi=0$, and Eq. (11) leads to the relation $\psi=-3r_+\psi'/2$.
Asymptotic solutions to Eqs. (10) and (11) are given by
\begin{equation}
\phi(r)=\mu-\frac{\rho}{r}+\cdot\cdot\cdot,
 \label{12}
\end{equation}
\begin{equation}
\psi=\frac{\psi^{(1)}}{r}+\frac{\psi^{(2)}}{r^2}+\cdot\cdot\cdot.
 \label{13}
\end{equation}
Parameters $\mu$ and $\rho$ are the chemical potential and the charge density of the dual field theory.
To have the asymptotic AdS$_4$ region to be stable one should use $\psi^{(1)}\neq 0$, $\psi^{(2)}=0$ or $\psi^{(1)}=0$, $\psi^{(2)}\neq0$.
In the following we set $\psi^{(2)}=0$ and $\psi^{(1)}\neq 0$. The field $\psi$ defines the condensate of the scalar operator ${\cal O}$ in the dual theory \cite{Hartnoll} so that $\psi^{(i)}=\langle{\cal O}_i\rangle/\sqrt{2}$ ($i=1,2$).

\section{Critical temperature of holographic superconductor}

Let us study the dependence of the critical temperature on the charge density for the holographic superconductor with arcsin-electrodynamics.
The real and positive solution to Eq. (10) at $\psi=0$ is
\begin{equation}
\phi'(r)=\frac{\sqrt{2}}{C\beta}\sqrt{\sqrt{r^8+C^4\beta^2}-r^4},
 \label{14}
\end{equation}
where $C$ is a dimensionless constant of integration. It follows from Eq. (14) that the electric field $E=-\phi'(r)$ at the origin is finite \cite{Kruglov} $E(0)=\sqrt{2/\beta}$. The same attractive feature occurs in BI nonlinear electrodynamics \cite{Born}. Expanding the function (14) in the small dimensionless parameter $C^4\beta^2/r^8$ we obtain
\begin{equation}
\phi'(r)=\frac{C}{r^2}-\frac{C^5\beta^2}{8r^{10}} +{\cal O}(\beta^4r^{-16}).
 \label{15}
\end{equation}
Integrating Eq. (15) and introducing a new variable $z=r_+/r$, we obtain ($C=\rho$) the approximate value
\begin{equation}
\phi(z)=\int_{r_{+(c)}}^r \phi'(r)dr=\frac{\rho}{r_{+(c)}}(1-z)+\frac{\rho^5\beta^2(z^9-1)}{72r_{+(c)}^{9}}.
 \label{16}
\end{equation}
Because solution (16) corresponds to $T=T_c$ ($\psi=0$) we use in Eq. (16) $r_+=r_{+(c)}$.
Thus, the corrections to Maxwell's electrodynamics in Eq. (16) are in the order of ${\cal O}(\beta^2/r^9)$.  The value $z = 1$ corresponds to the horizon $r=r_+$ and $z = 0$ corresponds to the boundary $r\rightarrow \infty$. The metric function is $f(z)=r_+^2(1-z^3)/z^2$.
In terms of the variable $z$ Eq. (11) is given by
\begin{equation}
\psi''(z)-\frac{2+z^3}{z(1-z^3)}\psi'(z)+\left[\frac{\phi^2(z)}{r_+^2(1-z^3)^2}+\frac{2}{z^2(1-z^3)}\right]\psi(z)=0.
 \label{17}
\end{equation}
From Eq. (16), up to ${\cal O}(\beta^2)$, one finds
\begin{equation}
\phi^2(z)\approx \lambda^2r_{+(c)}^2(z-1)^2\left(1-\frac{\lambda^4\beta^2\xi(z)}{36}\right),
 \label{18}
\end{equation}
where $\lambda=\rho/r_{+(c)}^2$ and $\xi(z)=(z^9-1)/(z-1)$. Making use of Eq. (18), Eq. (17) near the critical point $T\approx T_c$ takes the form
\[
\psi''(z)-\frac{2+z^3}{z(1-z^3)}\psi'(z)+\frac{2}{z^2(1-z^3)}\psi(z)
\]
\begin{equation}
+\frac{\lambda^2}{(1+z+z^2)^2}\left[1-\frac{\lambda^4\beta^2\xi(z)}{36}\right]\psi(z)=0.
 \label{19}
\end{equation}
At the boundary we define the scalar field $\psi(z)$ as
\begin{equation}
\psi(z)=\frac{\langle{\cal O}_1\rangle}{\sqrt{2}r_+}zF(z),
 \label{20}
\end{equation}
with the trial function $F(z)=1-\alpha z^2$ which satisfies, near the boundary, the conditions $F(0)=1$, $F'(0)=0$ \cite{Siopsis}.
Putting Eq. (20) into Eq. (19) we obtain the differential equation for the trial function
\begin{equation}
\left[(1-z^3)F'(z)\right]'-zF(z)+\frac{\lambda^2(1-z)}{1+z+z^2}\left[1-\frac{\lambda^4\beta^2\xi(z)}{36}\right]F(z)=0.
 \label{21}
\end{equation}
Comparing Eq. (21) with the Sturm-Liouville form
\begin{equation}
\left[p(z)F'(z)\right]'+q(z)F(z)+\lambda^2g(z)F(z)=0,
 \label{22}
\end{equation}
we obtain functions as follows:
\begin{equation}
p(z)=1-z^3,~~q(z)=-z,~~g(z)=\frac{(1-z)}{1+z+z^2}\left[1-\frac{\lambda^4\beta^2\xi(z)}{36}\right].
 \label{23}
\end{equation}
Now we are in the position to estimate the minimum eigenvalue of $\lambda^2$ \cite{Gangopadhyay}
\begin{equation}
\lambda^2=\frac{\int_0^1\{p(z)[F'(z)]^2-q(z)[F(z)]^2\}dz}{\int_0^1\{g(z)[F(z)]^2\}dz}.
 \label{24}
\end{equation}
For the function $g(z)$ in Eq. (23) we can retain terms which are linear in the parameter $\beta^2$ so that $\beta^2\lambda^4=\beta^2\lambda^4|_{\beta=0}+{\cal O}(\beta^4)$. Thus, we will use the value of $\lambda^4$ in Eq. (23) for $\beta^2=0$.
Putting $\beta=0$ in Eq. (23) one obtains from Eq. (24) the expression as follows (see also \cite{Sheykhi}):
\begin{equation}
\lambda^2|_{\beta=0}=\frac{2(5\alpha^2-3\alpha+3)}{6(2\alpha^2+2\alpha-1)\ln(3)-13\alpha^2-36\alpha+2\sqrt{3}\pi(2\alpha+1)}.
 \label{25}
\end{equation}
The critical temperature (3) (for $L=1$ and $M=r_+^3$) in terms of $\lambda=\rho/r^2_{+(c)}$ reads
\begin{equation}
T_c=\frac{3}{4\pi}\sqrt{\frac{\rho}{\lambda}}.
 \label{26}
\end{equation}
Minimizing the value $\lambda^2$, Eq. (24), with respect to $\alpha$ we find, according to Eq. (26), the maximum value of $T_c$. Then one can obtain the critical temperature from Eq. (26).

Critical temperatures for different parameters $\beta^2$ are given in Table 1.
\begin{table}[ht]
\caption{Critical temperature}
\centering
\begin{tabular}{c c c c c c c c}\\[1ex]
 \hline
$\beta^2$ & 0 & 0.1 & 0.2 & 0.3 & 0.4 & 0.5 & 0.6 \\[1ex]
\hline 
 $\alpha$ & 0.2389 & 0.2392 & 0.2395 & 0.2399 & 0.2401 & 0.2404 & 0.2407 \\[1ex]
\hline
 $\lambda^2_{min}$ & 1.2683 & 1.2770 & 1.2858 & 1.2948 & 1.3038 & 1.3130 & 1.3223 \\[1ex]
\hline
 $T_c/\sqrt{\rho}$ & 0.2250 & 0.2246 & 0.2242 & 0.2238 & 0.2234 & 0.2230 & 0.2226 \\[1ex]
\hline
\end{tabular}
\end{table}
In accordance with Table 1, when the parameter $\beta$ increases the critical temperature decreases. For definite parameter $\beta$ the critical temperature for BI electrodynamics and exponential electrodynamics (EN) (at $b=\beta$) is smaller compared to arcsin-electrodynamics. As a result, in our case of arcsin-electrodynamics, to create the condensation is easier comparing to BI and EN electrodynamics but more harder with respect to classical Maxwells's electrodynamics.

\section{Condensates and the critical exponent}

To calculate condensates in the dual theory we consider Eq. (10) near the critical point. Equation (10) in terms of the variable $z=r_+/r$ is written as
\begin{equation}
\left(1+\frac{\beta^2z^8}{4r_+^4}\phi^{'4}(z)\right)\phi''(z)+\beta^2z^7\frac{\phi^{'5}(z)}{r_+^4}
-\frac{2\phi(z)\psi^2(z)}{z^2(1-z^3)}\left(1-\frac{\beta^2z^8}{4r_+^4}\phi^{'4}(z)\right)^{3/2}=0.
 \label{27}
\end{equation}
Replacing Eq. (20) into Eq. (27), one obtains
\[
\left(1+\frac{\beta^2z^8}{4r_+^4}\phi^{'4}(z)\right)\phi''(z)+\beta^2z^7\frac{\phi^{'5}(z)}{r_+^4}
\]
\begin{equation}
=\frac{\phi(z)\langle{\cal O}_1\rangle^2F^2(z)}{r_+^2(1-z^3)}\left(1-\frac{\beta^2z^8}{4r_+^4}\phi^{'4}(z)\right)^{3/2}.
 \label{28}
\end{equation}
Assuming that $\langle{\cal O}_1\rangle^2/r_+^2$ is a small parameter \cite{Banerjee} we expand $\phi(z)$ as
\begin{equation}
\frac{\phi(z)}{r_+}=\frac{\phi_0(z)}{r_+}+\frac{\langle{\cal O}_1\rangle^2}{r_+^2}\chi(z)+\cdot\cdot\cdot,
 \label{29}
\end{equation}
where $\phi_0(z)$ satisfies  Eq. (28) for $\langle{\cal O}_1\rangle=0$,
\begin{equation}
\left(1+\frac{\beta^2z^8}{4r_+^4}\phi^{'4}_0(z)\right)\phi''_0(z)+\beta^2z^7\frac{\phi^{'5}_0(z)}{r_+^4}=0.
 \label{30}
\end{equation}
Inserting (29) into Eq. (28) and neglecting small parameters, we find the differential equation for the function $\chi(z)$:
\[
\frac{d}{dz}\left[\chi'(z)\left(1+\frac{\beta^2z^8}{4r_+^4}\phi^{'4}_0(z)\right)\right]+3\beta^2z^7\frac{\phi^{'4}_0(z)}{r_+^4}\chi'(z)
\]
\begin{equation}
=\frac{\phi_0(z)F^2(z)}{r_+(1-z^3)}\left(1-\frac{\beta^2z^8}{4r_+^4}\phi_0^{'4}(z)\right)^{3/2}.
 \label{31}
\end{equation}
The function $\chi(z)$ obeys the boundary conditions $\chi(1)=\chi'(1)=0$. The approximate solution to Eq. (30) is given by
\begin{equation}
\phi_0(z)=\frac{\rho(1-z)}{r_{+(c)}} +\frac{\rho^5\beta^2(z^9-1)}{72r_{+(c)}^9}+{\cal O}(\beta^4).
 \label{32}
\end{equation}
Substituting Eq. (32) into Eq. (31), we find up to ${\cal O}(\beta^2)$ the equation as follows:
\[
\frac{d}{dz}\left[\chi'(z)\left(1+\frac{\beta^2\lambda^4z^8}{4}\right)\right]+3\beta^2\lambda^4z^7\chi'(z)
\]
\begin{equation}
+\frac{F^2(z)}{1+z+z^2}\left(\frac{\beta^2\lambda^5(\xi(z)+27z^8)}{72}-\lambda\right)=0,
 \label{33}
\end{equation}
where $\lambda=\rho/r_{+(c)}^2$ and $\xi(z)=(z^9-1)/(z-1)$. Equation (33) can be represented in the standard form
\begin{equation}
\zeta'(z)+P(z)\zeta(z)=Q(z),
 \label{34}
\end{equation}
where
\[
\zeta(z)=\chi'(z)\left(1+\frac{\beta^2\lambda^4z^8}{4}\right),~~P(z)=3\beta^2\lambda^4z^7,
\]
\begin{equation}
Q(z)=\frac{F^2(z)}{1+z+z^2}\left(\lambda-\frac{\beta^2\lambda^5(\xi(z)+27z^8)}{72}\right),
 \label{35}
\end{equation}
and we neglected terms containing $\beta^4$. The solution to Eq. (34) is well known and it is given by
\begin{equation}
\zeta(z)=\exp(-I(z))\left(\int Q(z)\exp(I(z))dz+C\right),
 \label{36}
\end{equation}
where $C$ is the constant of integration and
\begin{equation}
I(z)=\int P(z)dz=\frac{3}{8}\beta^2\lambda^4z^8.
 \label{37}
\end{equation}
Expanding the exponential function in small parameter $\beta^2$, we obtain
\begin{equation}
\int Q(z)\exp(I(z))dz=\frac{\lambda}{72}\int\frac{(1-\alpha z^2)^2}{1+z+z^2}\left(72-\beta^2\lambda^4\xi(z)\right)dz+{\cal O}(\beta^4),
\label{38}
\end{equation}
where we have used $F(z)=1-\alpha z^2$. The integral in Eq. (38) is complicated to write down. Now we use the boundary condition
$\chi'(1)=0$ that gives $\zeta(1)=0$. Then with the help of Eqs. (36) and (38) and the condition $\zeta(1)=0$, up to ${\cal O}(\beta^2)$, we obtain the constant of integration
\[
C=\frac{\lambda}{72}\biggl[12\alpha^2+ 144\alpha+24\sqrt{3}(\alpha^2-2\alpha-2)\arctan\sqrt{3}-36\alpha(\alpha+2)\ln(3)
\]
\begin{equation}
+\lambda^4\beta^2\left(\frac{183}{440}\alpha^2-\frac{11}{9}\alpha+\frac{39}{28}\right)\biggr].
 \label{39}
\end{equation}
Making use of Eqs. (36) and (39) one finds
\begin{equation}
\chi'(0)=\zeta(0)=-\lambda {\cal A},
 \label{40}
 \end{equation}
 \[
{\cal A}=-\frac{1}{72}\biggl[12\alpha^2+ 144\alpha+24\sqrt{3}(\alpha^2-2\alpha-2)\left(\arctan\sqrt{3}-\arctan\frac{1}{\sqrt{3}}\right)
\]
 \begin{equation}
-36\alpha(\alpha+2)\ln(3)+\lambda^4\beta^2\left(\frac{183}{440}\alpha^2-\frac{11}{9}\alpha+\frac{39}{28}\right)\biggr].
 \label{41}
 \end{equation}
 Now we explore the procedure described in \cite{Roychowdhury}. Comparing Eq. (12) with Eq. (29), taking into account Eqs. (26), (32) and (40),
 $T=3r_+/(4\pi)$, and using series $\chi(z)=\chi(0)+z\chi'(0)+\cdot\cdot\cdot$, we obtain the order parameter
\begin{equation}
\langle{\cal O}_1\rangle=\gamma T_c\sqrt{1-\frac{T}{T_c}},
 \label{42}
 \end{equation}
 \begin{equation}
\gamma=\frac{4\pi\sqrt{2}}{3\sqrt{{\cal A}}}.
 \label{43}
 \end{equation}
The condensation values $\gamma$ for different parameters $\beta^2$ are given in Table 2.
\begin{table}[ht]
\caption{Condensation values $\gamma$}
\centering
\begin{tabular}{c c c c c c c c}\\[1ex]
 \hline
$\beta^2$ & 0 & 0.1 & 0.2 & 0.3 & 0.4 & 0.5 & 0.6 \\[1ex]
\hline 
 $\alpha$ & 0.2389 & 0.2392 & 0.2395 & 0.2399 & 0.2401 & 0.2404 & 0.2407 \\[1ex]
\hline
 $\gamma$ & 8.074 & 8.281 & 8.303 & 8.327 & 8.350 & 8.375 & 8.400 \\[1ex]
\hline
\end{tabular}
\end{table}

\section{Condensation values with matching method}

Making use of the matching method \cite{Soda} we will derive analytic expressions
for the condensation value and the critical temperature. By virtue of Taylor series, near the horizon, one obtains
\begin{equation}
\phi(z)=\phi(1)+\phi'(1)(z-1)+\frac{1}{2}\phi''(1)(z-1)^2,
 \label{44}
 \end{equation}
 \begin{equation}
\psi(z)=\psi(1)+\psi'(1)(z-1)+\frac{1}{2}\psi''(1)(z-1)^2.
 \label{45}
 \end{equation}
From Eq. (27) we find the equation as follows:
\begin{equation}
\phi''(z)+\frac{\beta^2z^7\phi^{'5}(z)}{r_+^4}
-\frac{2\phi(z)\psi^2(z)}{z^2(1-z^3)}\left(1-\frac{5\beta^2z^8\phi^{'4}(z)}{8r_+^4}\right)+{\cal O}(\beta^4)=0.
 \label{46}
\end{equation}
Then, with the help of boundary condition $\phi(1)=0$, near $z = 1$, and from Eq. (44) we obtain
\begin{equation}
\phi''(1)=-\frac{\beta^2\phi^{'5}(1)}{r_+^4}
-\frac{2\phi'(1)\psi^2(1)}{3}\left(1-\frac{5\beta^2\phi^{'4}(1)}{8r_+^4}\right)+{\cal O}(\beta^4).
 \label{47}
 \end{equation}
Substituting Eq. (47) into Eq. (44) one finds
\[
\phi(z)=\phi'(1)(z-1)-\frac{1}{2}(z-1)^2\left[\frac{\beta^2\phi^{'5}(1)}{r_+^4}
+\frac{2\phi'(1)\psi^2(1)}{3}\left(1-\frac{5\beta^2\phi^{'4}(1)}{8r_+^4}\right)\right]
\]
\begin{equation}
+{\cal O}(\beta^4).
 \label{48}
 \end{equation}
From Eqs. (17) and (45) by taking into account the boundary condition $\psi'(1)=2\psi(1)/3$, near $z = 1$, we obtain
\begin{equation}
\psi''(1)=-\frac{\phi^{'2}(1)\psi(1)}{9r_+^2}.
 \label{49}
 \end{equation}
Placing (49) into (45) with $\psi'(1)=2\psi(1)/3$ one finds
\begin{equation}
\psi(z)=\psi(1)+\frac{2}{3}\psi(1)(z-1)-\frac{\phi^{'2}(1)\psi(1)}{18r_+^2}(z-1)^2.
 \label{50}
 \end{equation}
Taking into consideration the method of \cite{Soda} (see also \cite{Roychowdhury_0}), we will obtain an analytic expression for the
critical temperature $T_c$. Matching asymptotic solutions (12), (13) with (48) and (50) at  $z = z_m$ we obtain
\begin{equation}
\mu-\frac{\rho z_m}{r_+}=\phi'(1)(z_m-1)-\frac{1}{2}(z_m-1)^2\phi'(1)\left[\frac{\beta^2\phi^{'4}(1)}{r_+^4}
+\frac{2\psi^2(1)}{3}\left(1-\frac{5\beta^2\phi^{'4}(1)}{8r_+^4}\right)\right],
 \label{51}
 \end{equation}
\begin{equation}
\frac{\psi^{(1)}z_m}{r_+}=\frac{1}{3}\psi(1)+\frac{2}{3}\psi(1)z_m-\frac{\phi^{'2}(1)\psi(1)}{18r_+^2}(z_m-1)^2.
 \label{52}
 \end{equation}
To match these asymptotic solutions smoothly one needs the equality also of derivatives at $z = z_m$
\begin{equation}
\frac{\rho}{r_+}=-\phi'(1)+(z_m-1)\phi'(1)\left[\frac{\beta^2\phi^{'4}(1)}{r_+^4}
+\frac{2\psi^2(1)}{3}\left(1-\frac{5\beta^2\phi^{'4}(1)}{8r_+^4}\right)\right],
 \label{53}
 \end{equation}
\begin{equation}
\frac{\psi^{(1)}}{r_+}=\frac{2}{3}\psi(1)-\frac{\phi^{'2}(1)\psi(1)}{9r_+^2}(z_m-1).
 \label{54}
 \end{equation}
With the help of notations $b =-\phi'(1)$, $a = \psi(1)$ and Eq. (53) we obtain
\begin{equation}
a^2=\frac{3}{2(1-z_m)}\left[\frac{\rho}{br_+}-1+\beta^2\frac{b^4}{r_+^4}\left(\frac{5\rho}{8br_+}-\frac{13}{8}+z_m\right)\right]+{\cal O}(\beta^4).
 \label{55}
 \end{equation}
Making use of $T\equiv T_H=3r_+/(4\pi)$, from (55) we find
\begin{equation}
a^2=\frac{3}{2(1-z_m)}\left(\frac{T_c}{T}\right)^2\left(1+\frac{\beta^2\tilde{b}^4(13-8z_m)}{8}\right)\left(1-\frac{T^2}{T_c^2}\right)+{\cal O}(\beta^4),
 \label{56}
 \end{equation}
where $\tilde{b}=b/r_+$ and
\begin{equation}
T_c=\frac{3\sqrt{\rho}}{4\pi\sqrt{\tilde{b}}}\sqrt{1-\tilde{b}^4\beta^2(1-z_m)}.
 \label{57}
 \end{equation}
Close to the critical temperature ($T \approx T_c$), from (56) we obtain
\begin{equation}
a=\sqrt{\frac{3}{1-z_m}}\left(1+\frac{\beta^2\tilde{b}^4(13-8z_m)}{16}\right)\sqrt{1-\frac{T}{T_c}}+{\cal O}(\beta^4).
 \label{58}
 \end{equation}
From (52) and (54) one can find
\begin{equation}
\tilde{b}=\sqrt{\frac{6}{1-z_m^2}},~~~~\psi^{(1)}=\frac{2ar_+(2+z_m)}{3(1+z_m)}.
 \label{59}
 \end{equation}
The condensation value $\langle{\cal O}_1\rangle=\sqrt{2}\psi^{(1)}$, near the critical temperature ($T \approx T_c$) obtained from Eqs, (58) and (59), is given by
\[
\langle{\cal O}_1\rangle=\frac{8\sqrt{2}\pi}{9}\left(\frac{2+z_m}{1+z_m}\right)\sqrt{\frac{3}{1-z_m}}
\left(1+\frac{\beta^2\tilde{b}^4(13-8z_m)}{16}\right)
\]
\begin{equation}
\times T_c\sqrt{1-\frac{T}{T_c}}+{\cal O}(\beta^4).
 \label{60}
 \end{equation}
The similar dependence of condensates on the temperature holds in the Sturm-Liouville eigenvalue problem method.
It follows from Eq. (57), for the critical temperature ($T_c$,) that there is the following upper bound on the coupling parameter $\beta$
\begin{equation}
\beta^2\leq \frac{(1-z_m^2)^2}{36(1-z_m)}.
 \label{61}
 \end{equation}
Thus, the presence of nonlinear corrections makes the critical temperature lower, and therefore, it is harder to have the
scalar condensate at low temperature.
In Fig. 1 the critical temperature $T_c$ vs. $\rho$ for s-wave holographic superconductors with
$z_m = 1/2$ for different choices of the parameter $\beta^2$ is depicted.
\begin{figure}[h]
\includegraphics[height=3.0in,width=3.0in]{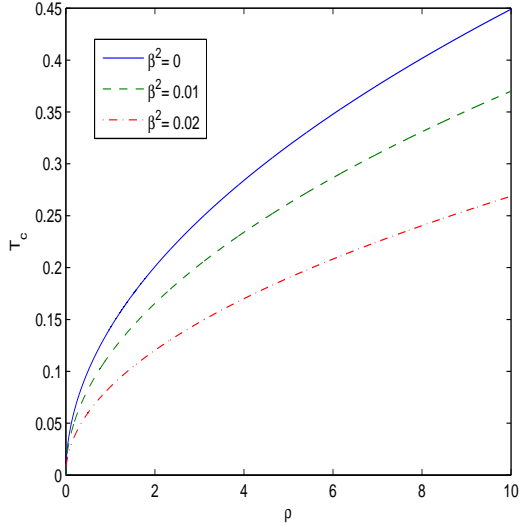}
\caption{\label{fig.1}The plot of critical temperature $T_c$ vs. $\rho$ for different choices of the parameter $\beta^2$.}
\end{figure}
The plot of $T_c/\sqrt{\rho}$ vs. the matching parameter $z_m$ is given in Fig. 2.
\begin{figure}[h]
\includegraphics[height=3.0in,width=3.0in]{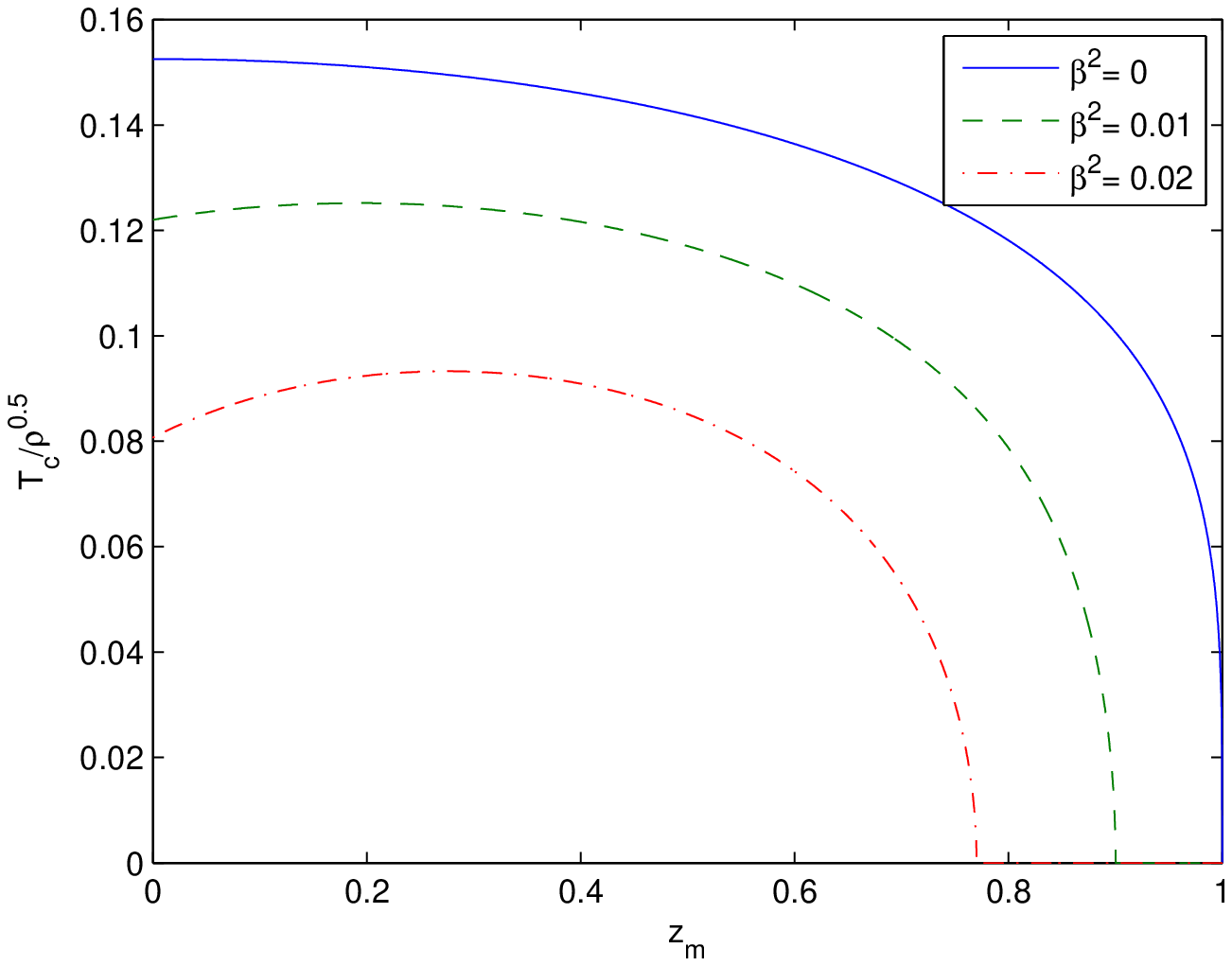}
\caption{\label{fig.2}The $T_c/\sqrt{\rho}$ vs. the matching parameter $z_m$.}
\end{figure}
The plot of the condensation values $\langle{\cal O}_1\rangle/T_c$ vs. $T/T_c$ ($z_m=0.5$) is represented in Fig. 3.
\begin{figure}[h]
\includegraphics[height=3.0in,width=3.0in]{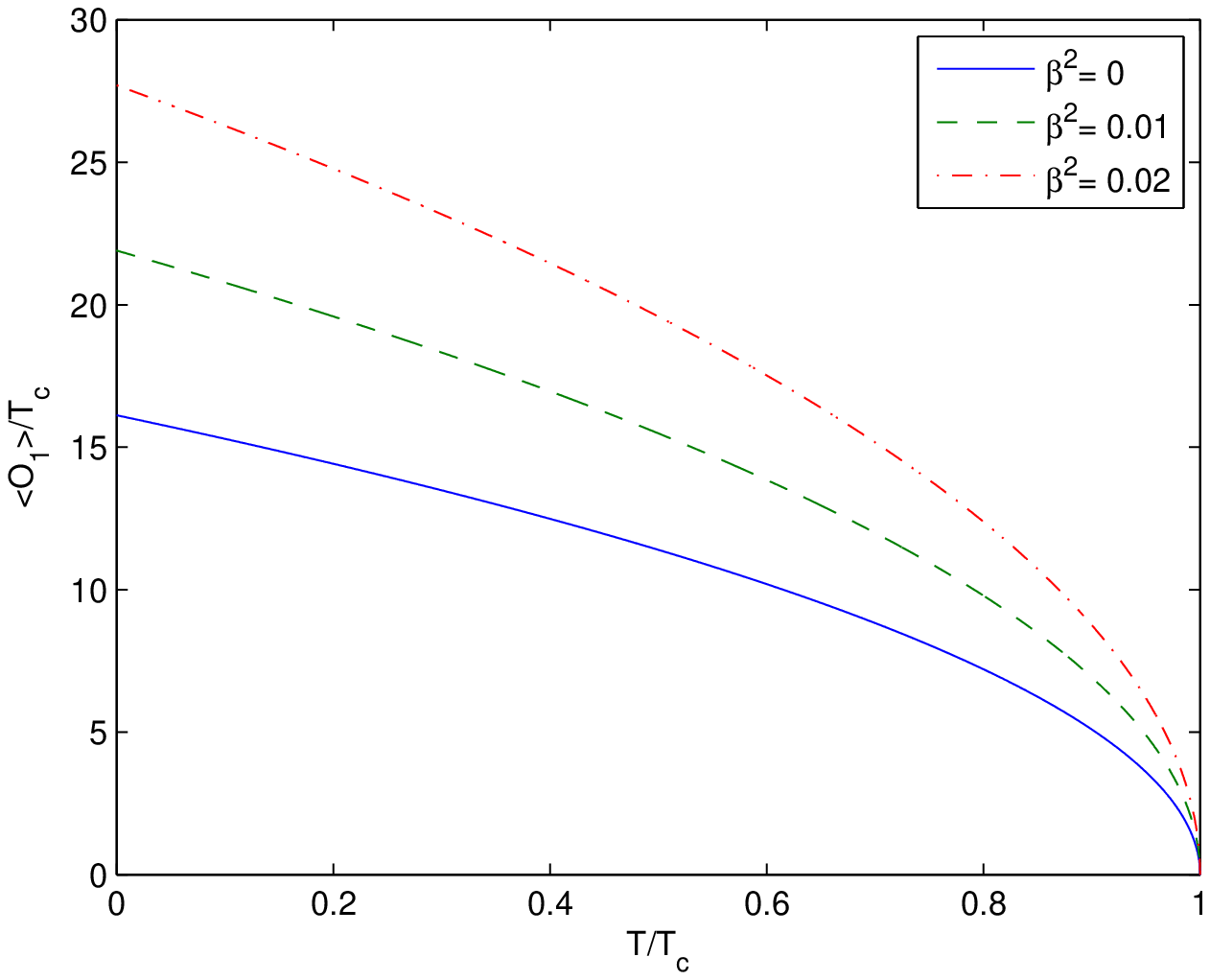}
\caption{\label{fig.3}The condensation values $\langle{\cal O}_1\rangle/T_c$ vs. $T/T_c$ .}
\end{figure}
It follows from Fig. 2 that the range for the matching parameter $z_m$ decreases when the parameter $\beta$ increases. However, the different choice of $z_m$ (close to $z_m=0.5$)) does not give a big difference in numerical values for the critical temperature.

Let us compare the values for the critical temperature calculated from the method based on the Sturm-Liouville eigenvalue problem and from the matching method for $\beta=0.01$. From the Sturm-Liouville eigenvalue problem we find $T_c/\rho \approx 0.225$ and from the matching method ($z_m=0.5$) one obtains $T_c/\rho \approx 0.117$. Thus, there is a big discrepancy in values for the critical temperature in both methods. The same situation occurs for condensates computed in these methods. But the method based on the Sturm-Liouville eigenvalue problem gives more precise values close to the one calculated from the numerical method. In addition, the matching method  allows to consider only very small values of the parameter $\beta$ due to the restriction from Eq. (61).

\section{Conductivity}

Next, we evaluate the low temperature conductivity in the dual theory. To find the conductivity on the $AdS_4$ boundary, one needs to obtain the  vector potential $A_x(t,r)$ by applying an electromagnetic perturbation. From Eq. (7) we find the equation along the boundary as follows (for $q=1$ and $\psi$ is real):
\begin{equation}
\frac{1}{r^2}\partial_0\left(\frac{r^2F^{0x}}{\sqrt{1-(\beta{\cal F})^2}}\right)+\frac{1}{r^2}\partial_r\left(\frac{r^2F^{rx}}{\sqrt{1-(\beta{\cal F})^2}}\right)=2A^x\psi^2,
 \label{62}
\end{equation}
where ${\cal F}=-E_xE^x/2=-(\partial_0A_x)\partial_0A^x/2$. Making use of the sinusoidal electromagnetic perturbation $A_x$ of the form $\exp (-i\omega t)$ and relations $A^x=A_x/r^2$, $F^{0x}=-\partial_0 A_x/(f(r)r^2)$, $F^{rx}=f(r)(\partial_r A_x)/r^2$, we obtain from Eq. (62)
\begin{equation}
\partial_0\left(\frac{i\omega A_x}{f(r)\sqrt{1-\beta^2\omega^4A^4/4}}\right)
+\partial_r\left(\frac{f(r)\partial_rA_x}{\sqrt{1-\beta^2\omega^4A^4/4}}\right)=2A_x\psi^2,
 \label{63}
\end{equation}
where $A^2=A_xA^x$. From Eq. (63) one finds
\[
\frac{\omega^2A_x}{f(r)}+\frac{\omega^6\beta^2 A_xA^4}{2f(r)(1-\beta^2\omega^4A^4/4)}+\left(f(r)A'_x\right)'
+\frac{\omega^4\beta^2f(r)A'_x(A^4)'}{8(1-\beta^2\omega^4A^4/4)}
\]
\begin{equation}
=2A_x\psi^2\sqrt{1-\beta^2\omega^4A^4/4},
 \label{64}
\end{equation}
where the prime means the partial derivative on the $r$ ($A'_x =\partial_r A_x$). Making use of Taylor series in small parameter $\beta^2$ we obtain the equation as follows:
\[
\left(f(r)A'_x\right)'+\left(\frac{\omega^2}{f(r)}-2\psi^2\right)A_x
\]
\begin{equation}
=\frac{\beta^2\omega^4}{4}\left(-A_xA^4\psi^2-\frac{2\omega^2 A_xA^4}{f(r)}-\frac{f(r)A'_x\left(A^4\right)'}{2} \right)+{\cal O}(\beta^4).
 \label{65}
\end{equation}
At $\beta=0$ one arrives at the equation found in \cite{Hartnoll}. We now calculate corrections to the conductivity due to nonlinear terms of arcsin-electrodynamics. We look for a solution to Eq. (65), up to ${\cal O}(\beta^2)$, in the form
\begin{equation}
A_x= \bar{A_x}+\beta^2B,
 \label{66}
\end{equation}
where $\bar{A_x}$ obeys the homogenous equation
 \begin{equation}
\left(f(r)\bar{A'_x}\right)'+\left(\frac{\omega^2}{f(r)}-2\psi^2\right)\bar{A_x}=0.
 \label{67}
\end{equation}
From Eqs. (65) and (66) we obtain the equation
\[
\left(f(r)B'\right)'+\left(\frac{\omega^2}{f(r)}-2\psi^2\right)B
\]
\begin{equation}
=\frac{\omega^4}{4}\left(-\bar{A}_x\bar{A}^4\psi^2-2\frac{\omega^2 \bar{A}_x\bar{A}^4}{f(r)}-\frac{f(r)\bar{A}'_x\left(\bar{A}^4\right)'}{2} \right)+{\cal O}(\beta^4).
 \label{68}
\end{equation}
Near the boundary $(r\rightarrow \infty$) $f(r)\approx r^2$, $\psi\approx \langle {\cal O}_1\rangle/(\sqrt{2}r)$ and the solution to Eq. (67) is given by \cite{Hartnoll}
 \begin{equation}
\bar{A_x}=\exp\left( i\frac{\sigma_0\omega}{r}\right),
 \label{69}
\end{equation}
where the conductivity for $\beta=0$ is
\begin{equation}
\sigma_0=\sqrt{1-\frac{\langle {\cal O}_1\rangle^2}{\omega^2}}.
 \label{70}
\end{equation}
This solution corresponds to low temperatures and the limit $M\rightarrow 0$.
Inserting Eq. (69) into Eq. (68), and making use of Eq. (70), we obtain
\[
\left(f(r)B'\right)'+\left(\frac{\omega^2}{f(r)}-2\psi^2\right)B
\]
\begin{equation}
=-\frac{\omega^4}{4}\left[\left(\frac{\langle {\cal O}_1\rangle^2}{2}+2\omega^2-2\sigma_0\omega^2\right)\frac{1}{r^6}+ \frac{2i\sigma_0\omega}{r^5}\right]\exp\left(\frac{5i\sigma_0\omega}{r}\right)+{\cal O}(\beta^4).
 \label{71}
\end{equation}
We find the solution of Eq. (71) in the form
\begin{equation}
B=\left(c_0+\frac{c_1}{r}+\frac{c_2}{r^2}+\frac{c_3}{r^3}+\frac{c_4}{r^4}\right)\exp\left(\frac{5i\sigma_0\omega}{r}\right)+{\cal O}(\beta^4),
 \label{72}
\end{equation}
where
\[
c_0=\frac{1415\sigma_0^2-2195}{165888\sigma_0^6},~~~
c_1=\frac{i\omega(625-397\sigma_0^2)}{13824\sigma_0^5},~~~
c_2=\frac{\omega^2(155-95\sigma_0^2)}{2304\sigma_0^4},
\]
\begin{equation}
c_3=\frac{i\omega^3(13\sigma_0^2-25)}{576\sigma_0^3},~~~
c_4=\frac{5\omega^4(1-\sigma_0^2)}{192\sigma_0^2}.
 \label{73}
\end{equation}
Making use of Eqs. (66), (69) and (72) at large radius, and neglecting terms ${\cal O}(\beta^4)$, one finds
\begin{equation}
A_x=1+\frac{i\sigma_0\omega}{r}+\beta^2\left(c_0+\frac{c_1}{r}+\frac{5i\sigma_0\omega c_0}{r}\right)+{\cal O}(r^{-2}).
 \label{74}
\end{equation}
With the help of equations \cite{Hartnoll}
\begin{equation}
\sigma(\omega)=-\frac{iA_x^{(1)}}{\omega A_x^{(0)}},~~~A_x=A_x^{(0)}+\frac{A_x^{(1)}}{r}+{\cal O}(r^{-2}),
\label{75}
\end{equation}
where in accordance with the AdS/CFT correspondence, $A_x^{(0)}$ is the source and $A_x^{(1)}$ is dual to the
current, and making use of Eqs. (73),(74) and (75), we obtain the conductivity
\[
\sigma(\omega)=\sigma_0+\beta^2\left(4\sigma_0c_0-\frac{ic_1}{\omega}\right)+{\cal O}(\beta^4)
\]
\begin{equation}
=\sigma_0+\beta^2\left(\frac{7\sigma_0^2-10}{1296\sigma_0^5}\right)+{\cal O}(\beta^4).
\label{76}
\end{equation}
It was shown in \cite{Hartnoll} that the conductivity $\sigma_0$ is in very good agreement with numerical results as in high frequencies as well as in low frequencies at $\omega\ll \langle {\cal O}_1\rangle$. Therefore, our analytical calculations also should be in
good agreement with numerical results. The conductivity (76) gives corrections to $\sigma_0$ in the order of ${\cal O}(\beta^2)$. These corrections are due to exploring arcsin-electrodynamics instead of Maxwell's electrodynamics. The attractive feature of arcsin-electrodynamics is that the electric field in the centre of particles as well as the total electrostatic energy are finite. At high frequencies $\omega>\langle {\cal O}_1\rangle$ the conductivity (76) is real, $\sigma=\mbox{Re}[\sigma]$. For low frequencies $\omega<\langle {\cal O}_1\rangle$ we find the imaginary part of the conductivity from Eq. (76)
\begin{equation}
\mbox{Im}[\sigma]=\sqrt{\frac{\langle {\cal O}_1\rangle^2}{\omega^2}-1}
+\beta^2\left(\frac{7\langle {\cal O}_1\rangle^2/\omega^2+3}{1296(\langle {\cal O}_1\rangle^2/\omega^2-1)^{5/2}}\right)+{\cal O}(\beta^4).
\label{77}
\end{equation}
The plots of $\mbox{Re}[\sigma]$ and $\mbox{Im}[\sigma]$ versus $\omega/\langle {\cal O}_1\rangle$ are depicted in figures 4 and 5.
\begin{figure}[h]
\includegraphics[height=3.0in,width=3.0in]{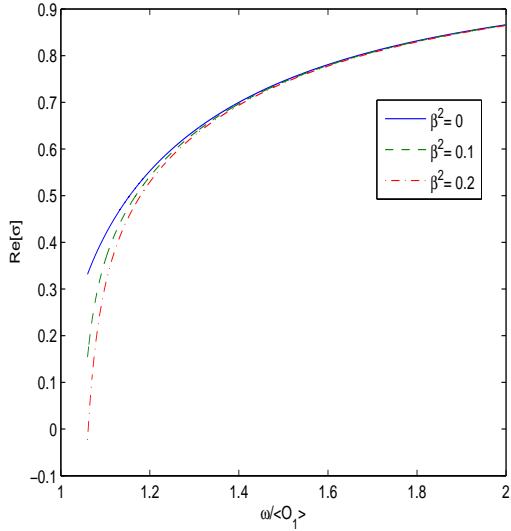}
\caption{\label{fig.4} The real part of the conductivity vs. $\omega/\langle{\cal O}_1\rangle$.}
\end{figure}
\begin{figure}[h]
\includegraphics[height=3.0in,width=3.0in]{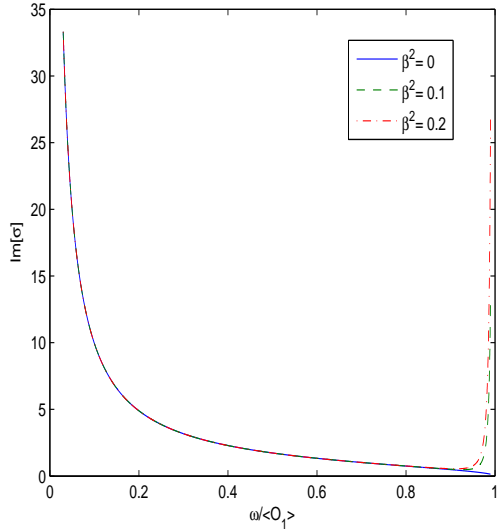}
\caption{\label{fig.5} The imaginary part of the conductivity vs. $\omega/\langle{\cal O}_1\rangle$.}
\end{figure}
According to figure 4 the corrections to Re$[\sigma_0]$ lowered the conductivity Re[$\sigma$]. But at $\omega\rightarrow \langle{\cal O}_1\rangle$ the Im[$\sigma$] has a pole and becomes infinite. This pole is nonphysical because we have implied that corrections to conductivity  $\sigma_0$ are small. Therefore corrections to conductivity  $\sigma_0$ around the point $\omega=\langle{\cal O}_1\rangle$ are beyond the method used.
As it follows from Eq. (77) the imaginary part of the conductivity at $\omega=0$ has a pole, Im$[\sigma(\omega)]=\langle{\cal O}_1\rangle/\omega$. It is interesting that corrections to the conductivity, due to nonlinear terms of arcsin-electrodynamics, do not contribute to the coefficient $\langle{\cal O}_1\rangle$ of this pole. Then the real part of the conductivity, making use of the Kramers-Kronig relations, possesses a delta function, Re$[\sigma(\omega)]=\pi \langle{\cal O}_1\rangle\delta(\omega)$ \cite{Hartnoll}. At $\omega=0$ the model possesses a DC superconductivity. The DC superconductivity is infinite because of a second order phase transition. The real part of the conductivity versus $\omega/T$ at $\beta^2=0.2$ is represented in Fiq. 6.
\begin{figure}[h]
\includegraphics[height=3.0in,width=3.0in]{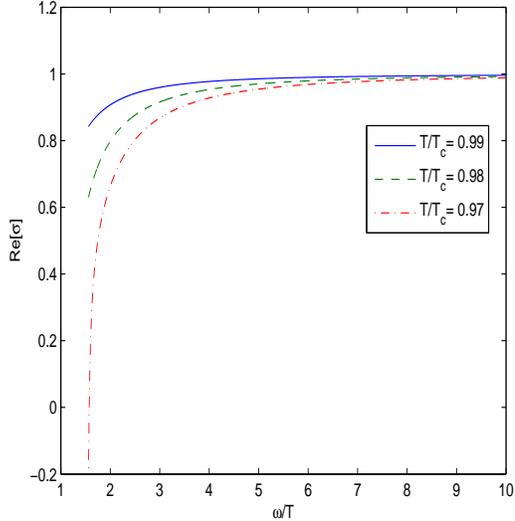}
\caption{\label{fig.6} The real part of the conductivity vs. $\omega/T$ at $\beta^2=0.2$.}
\end{figure}
Figure 6 shows the formation of a gap in the real part of the conductivity. At high frequencies we have the normal phase and there is no a condensate. If temperatures are low (and below the critical temperature) the gap opens.

 \section{Conclusion}

We have investigated the holographic s-wave superconductors in the background of Schwarzschild AdS$_4$ black holes when the gage field is described by arcsin-electrodynamics. The analytical approach based on the Sturm-Liouville eigenvalue problem was used. It should be noted that the numerical calculations are not accurate when the temperature approaches to zero \cite{Hartnoll2}, \cite{Siopsis}.
We used an approximation that the scalar field and electromagnetic fields, in the bulk theory, do not effect on the background (the probe limit). Near the critical point the condensation operator and the critical temperature depend on the parameter $\beta$ of the model of arcsin-electrodynamics. With increasing the parameter $\beta$  the critical temperature decreases. The similar property holds in the model of holographic s-wave superconductors with BI electrodynamics. The attractive feature of our model is that the condensation formation depends on $\beta$ weakly compared to models with BI and EN electrodynamics. The critical exponent of holographic s-wave superconductors is equal to $1/2$ as in other models. We have calculated the order parameter depending on $\beta$. Making use of the matching method we found analytic expressions
for the condensation value and the critical temperature. It should be noted that the critical temperature and the condensation values depend on the $z_m$ weakly. The method based on the Sturm-Liouville eigenvalue problem and the matching method, gave similar behaviour of the critical temperature and condensates. In both cases the critical exponent near the critical point is equal to $1/2$ in agreement with the value of the mean field approach. This indicates on the second-order phase transition near the critical temperature.
But numerical values found from the matching method are very approximate. In addition, the matching method leads to the upper bound on the parameter $\beta$ above which the analysis can not be carried out. We have computed  the conductivity in our model making use of an analytical method. Corrections to the real and imaginary parts of the conductivity were calculated up to ${\cal O}(\beta^2)$. It was shown that corrections to the conductivity $\sigma_0$, due to the presence of the parameter $\beta$, lowered $\sigma_0$.

\end{document}